\documentclass[preprint, superscriptaddress]{revtex4}%
\usepackage{graphicx}
\usepackage{bm}
\usepackage{epstopdf}
\usepackage{sidecap}
\usepackage{amsmath}
\usepackage{amsfonts}
\usepackage{amssymb}%
\providecommand{\U}[1]{\protect\rule{.1in}{.1in}}
\begin{document}
\title{Spatially coherent surface resonance states derived from magnetic resonances}
\author{Zeyong Wei}
\affiliation{Physics Department, Tongji University, Shanghai 200092, China}
\affiliation{Shanghai Key Laboratory of Special Artifical Microstructure Materials and
Technology, Shanghai, China}
\author{Hongqiang Li}
\email{hqlee@tongji.edu.cn}
\affiliation{Physics Department, Tongji University, Shanghai 200092, China}
\affiliation{Shanghai Key Laboratory of Special Artifical Microstructure Materials and
Technology, Shanghai, China}
\author{Yang Cao}
\affiliation{Physics Department, Tongji University, Shanghai 200092, China}
\affiliation{Shanghai Key Laboratory of Special Artifical Microstructure Materials and
Technology, Shanghai, China}
\author{Chao Wu}
\affiliation{Physics Department, Tongji University, Shanghai 200092, China}
\affiliation{Shanghai Key Laboratory of Special Artifical Microstructure Materials and
Technology, Shanghai, China}
\author{Jinzhi Ren}
\affiliation{Physics Department, Tongji University, Shanghai 200092, China}
\affiliation{Shanghai Key Laboratory of Special Artifical Microstructure Materials and
Technology, Shanghai, China}
\author{Zhihong Hang}
\affiliation{Department of Physics, Hong Kong University of Science and Technology, Clear
Water Bay, Kowloon, Hong Kong, China}
\author{Hong Chen}
\affiliation{Physics Department, Tongji University, Shanghai 200092, China}
\affiliation{Shanghai Key Laboratory of Special Artifical Microstructure Materials and
Technology, Shanghai, China}
\author{Daozhong Zhang}
\affiliation{Laboratory of Optical Physics, Institute of Physics, Chinese Academy of
Sciences, Beijing, China }
\author{C.T. Chan}
\affiliation{Department of Physics, Hong Kong University of Science and Technology, Clear
Water Bay, Kowloon, Hong Kong, China}

\begin{abstract}
A thin metamaterial slab comprising a dielectric spacer sandwiched between a
metallic grating and a ground plane is shown to possess spatially coherent
surface resonance states that span a large frequency range and can be tuned by
structural and material parameters. They give rise to nearly perfect
angle-selective absorption and thus exhibit directional thermal emissivity.
Direct numerical simulations show that the metamaterial slab supports
spatially coherent thermal emission in a wide frequency range that is robust
against structural disorder.

\end{abstract}

\maketitle

\section{Introduction}

Surface plasmon polaritons (SPPs) can modulate light waves at the
metal-dielectric interface with wavelength much smaller than that in free
space\cite{1},which enables the control of light in a subwavelength scale for
nanophotonic devices\cite{2}. SPPs with large coherent length are useful in
many areas, including optical processing, quantum information\cite{3} and
novel light-matter interactions\cite{4}. The enhancement of local fields by
SPPs is particularly important as it opens a new route to absorption
enhancement\cite{5}, nonlinear optical amplification\cite{6,7} as well as weak
signal probing\cite{8,9}. As the properties of SPP are pretty much determined
by the natural (plasmon) resonance frequency, there is not much room for us to
adjust the SPP response for practical applications. With induced surface
current oscillations on an array of metallic building
blocks\cite{10,11,12,13,14,15,16}, metamaterial surfaces can manipulate
electromagnetic waves in a similar way as SPPs. Such SPPs or surface resonance
states on structured metallic surfaces are tunable by geometric parameters.

In this paper, we examine the properties of surface resonance states at a
dielectric-metamaterial interface that exhibit magnetic response to the
incident waves and strong local field enhancement. We will see that these
surface resonance states can give highly directional absorptivity and
emissivity, and may thus help to realize interesting effects such as spatially
coherent thermal emission. As the structure is very simple, it can be
fabricated down to the IR and optical regime\cite{17,18,19}.

We will show that a thin metamaterial slab, with a thickness much smaller than
the operational wavelength, supports delocalized magnetic surface resonance
states with a long coherent length in a wide range of frequencies. Operating
in a broad frequency range, these spatially coherent SPPs are surface
resonance states with quasi-TEM modes guided in the dielectric layer that are
weakly coupled to free space, and the coupling strength can be controlled by
tuning structural parameters while the frequency can be controlled by varying
structural and material parameters. The high fidelity of these surface
resonance states results in directional absorptivity or emissivity, which is
angle-dependent with respect to frequency. Finite-difference-in-time-domain
(FDTD) simulations verify that the highly directional emissivity from the slab
persists in the presence of structural disorder in the grating layer.

Such metal-dielectric-metal (MDM) structures were recognized as artificial
magnetic surfaces with high impedance by the end of last century\cite{12} ,
the magnetic response were described with an effective permeability in Lorentz
type\cite{12,20}. After the concept of metamaterial being proposed\cite{21},
P. Alastair and his co-workers numerically and experimentally proved that the
ultra-thin MDM structures can resonantly absorb or transmit radiations at low
frequency limit[22]. They addressed that the central frequencies of absorption
peaks are independent from the incident angle with an interpretation of
Farby-Perrot resonant mode (EQ. 1 in Ref. 22). The same group further explored
the angle-independent absorption, as the main scenario of the incremental
work, by measuring the flat bands of surface wave dispersion in the
visible\cite{23} as well as the microwave region\cite{24}. In contrast, we
find that the structure with proper design also supports very narrow
absorption peaks which are sensitive to the incident angle and obviously do
not satisfy to the Fabry-Perot resonance condition suggested in the previous studies.

It is worth noting that the physics origin of an angle-independent peak is
quite different from that of angle-dependent ones. The former, investigated in
Refs. 22-24, mainly comes from the localized surface resonance states, while
the latter, found by us, comes from collective surface resonance states. An
intuitive picture is as follows: high order quasi-TEM modes induced inside the
dielectric layer can assign phase correlation to the outgoing waves emitted
from the air slits of grating, thus are very crucial to the formation of
collective response. Weak enough both the leakage from dielectric layer to air
slits and the material absorption, the spatial coherence of surface resonance
states will survive. As the interaction between the structure and the incident
waves will excite quasi-TEM modes inside the dielectric layer, the magnetic
induction must be parallel to the MDM surfaces if it exists. Thus a surface
resonance state on a MDM structure is usually magnetic in nature. Our findings
about spatially coherent surface resonance states are original compared to the
common knowledge, and have great potentials in coherent control of SPPs as
well as thermal emission radiations.

\section{Model Description and Mode Expansion Method}

Our model system is schematically illustrated in Fig. 1. Lying on the $\hat
{x}\hat{y}$ plane, the slab comprises an upper layer of a metallic lamellar
grating with thickness $t$, a dielectric spacer layer as a slab waveguide with
thickness $h$ and a metallic ground plane. The metallic strips are separated
by a small air gap $g$, giving rise to a period of $p=a+g$ for the lamellar
grating. Each metallic strip together with the ground plane beneath it
constitutes a planar resonant cavity as the building block that gives magnetic
responses at cavity resonances\cite{12,16} . The metallic grating is along the
$\hat{x}$ direction so that the guided waves in the dielectric layer (at
$0<z<h$ in region III) can only couple with the transverse magnetic ($TM)$
polarized wave (with the electric field$\vec{E}$in $\hat{x}\hat{z}$ plane) in
the free semi-space (at $z>h+t$ in region I). The geometric parameters of our
model are $t=0.2\mu$m, $h=0.8\mu$m, $a=3.8\mu$m, $g=0.2\mu$m and
$p=a+g=4.0\mu$m. \begin{figure}[ptb]
\centering
\includegraphics[width=12cm]{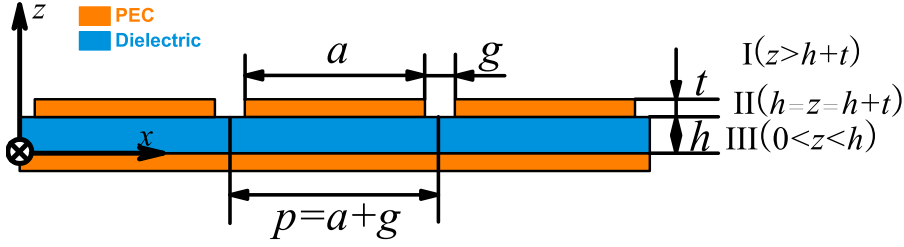}\caption{Schematic picture of the magnetic
metamaterial slab. The geometric parameters are $t=0.2\mu$m, $h=0.8\mu$m,
$a=3.8\mu$m, $g=0.2\mu$m and t $p=a+g=4.0\mu$m. the dielectric spacer layer is
slightly dissipative by assigning a complex permittivity$\varepsilon
_{III}=\varepsilon_{r}\varepsilon_{0}+i\frac{\sigma}{\omega}$ with
$\varepsilon_{r}=2.2$ and$\sigma=66.93$S/m.}%
\end{figure}For a TM incident plane wave with an in-plane wave vector $\vec
{k}_{0_{\parallel}}=k_{x}\hat{e}_{x}+k_{y}\hat{e}_{y}$, the EM field in region
I and in region III can be written in terms of the reflection coefficients
$r_{m}$ and the guided Bloch wave coefficients $t_{m}$\cite{25,26,27,28}, as
\begin{equation}%
\begin{array}
[c]{l}%
H_{1}(\vec{r},z)=\delta_{m,0}e^{-ik_{z_{m}}^{I}z}\langle\vec{r}|\vec
{k}_{0_{\parallel}}\rangle+\sum\limits_{m}{r_{m}e^{ik_{z_{m}}^{I}z}\langle
\vec{r}|\vec{k}_{m}^{I}\rangle}\\
H_{3}(\vec{r},z)=\sum\limits_{m}{[t_{m}e^{ik_{z_{m}}^{III}z}+t_{m}%
e^{-ik_{z_{m}}^{III}(z-2h)}]\langle\vec{r}|\vec{k}_{m}^{III}\rangle}%
\end{array}
\quad,\label{eq1}%
\end{equation}
where the term $\delta_{m,0}e^{-ik_{z_{m}}^{I}z}\langle\vec{r}|\vec
{k}_{0_{\parallel}}\rangle$ denotes the incident plane wave with $\langle
\vec{r}|\vec{k}_{0_{\parallel}}\rangle=e^{i(k_{x}x+k_{y}y)}$,$\delta_{m,0}$
being the Kronecker function and $m$ being the Bloch order; $\langle\vec
{r}|\vec{k}_{m}^{I}\rangle=\langle\vec{r}|\vec{k}_{m}^{III}\rangle=e^{i\vec
{k}_{m}\cdot\vec{r}}$denotes wave component of the $m^{th}$ Bloch eigenmode in
the semi-free space (region I) and the dielectric layer (region III) with
respect to $\vec{k}_{m}=\vec{k}_{0_{\parallel}}+\vec{G}_{m}$. $\vec{k}_{m}$is
the in-plane wave vector and $\vec{G}_{m}=\frac{2\pi m}{p}\hat{e}_{x}$ is the
$m^{th}$ reciprocal lattice vector. $k_{z_{m}}^{I}=\sqrt{\varepsilon_{0}%
\mu_{0}\omega^{2}-|\vec{k}_{m}|^{2}}$ and $k_{z_{m}}^{III}=\sqrt
{\varepsilon_{III}\mu_{0}\omega^{2}-|\vec{k}_{m}|^{2}}$ are the $z$ components
of wave vector for the $m^{th}$ order Bloch eigenmode in region I and region
III respectively. $\varepsilon_{0}\ $and $\varepsilon_{III}$ are the
permittivity of the vacuum and the dielectric slab, $\mu_{0}$ is the vacuum permeability.

We shall mainly consider infrared frequencies, at which the metals can be well
approximated as perfectly electric conductors (PEC). The EM fields at $h\leq
z\leq h+t$ in region II are squeezed inside the air gaps, in which the
magnetic fields can be expressed in terms of the expansion coefficients
$a_{l}$ and $b_{l}$ of forward and backward guided waves, as:
\begin{equation}
H_{2}(\vec{r},z)=\sum\limits_{l}{[a_{l}e^{-iq_{l}(z-h-t)}+b_{l}e^{iq_{l}%
(z-h)}]\langle\vec{r}}|\alpha_{l}\rangle\quad,\label{eq2}%
\end{equation}
where $\langle\hat{r}|\alpha_{l}\rangle=\cos[l\pi/g(x+g/2)]$,
$(l=0,1,...,n,...)$ is the in-plane distribution of guided mode $|\alpha
_{l}\rangle$ running over all air gaps defined by\cite{29}. $q_{l}%
=\sqrt{\varepsilon_{0}\mu_{0}\omega^{2}-(l\pi/g)^{2}-k_{y}^{2}}$ is the z
component of wave vector for the $l^{th}$ guided mode $|\alpha_{l}\rangle$.

We can obtain the coefficients $t_{m}(f,\vec{k}_{0_{//}})$ and $r_{m}%
(f,\vec{k}_{0_{//}})$ of the $m^{th}$ guided and reflected waves by applying
the boundary continuity conditions for the tangential components of
electromagnetic wave fields (over the slits) at the interfaces $z=h$ and
$z=h+t$. Given that surface resonance modes are intrinsic response, we can
also assign zero to the incident plane wave and apply the boundary continuity
conditions for the tangential components of wave fields to derive the
eigen-value equations. A surface resonance state can be determined by
searching a zero value/minimum of eigen-equation determinant in the reciprocal
space provided that it is non-radiative/radiative with infinite/finite life
time below/above light line in free space.

\section{ABSORPTION SPECTRA PROPERTIES AND SPACIAL COHERENCE OF MAGNETIC
SURFACE RESONACE STATES}

\begin{figure}[ptb]
\centering
\includegraphics[width=10cm]{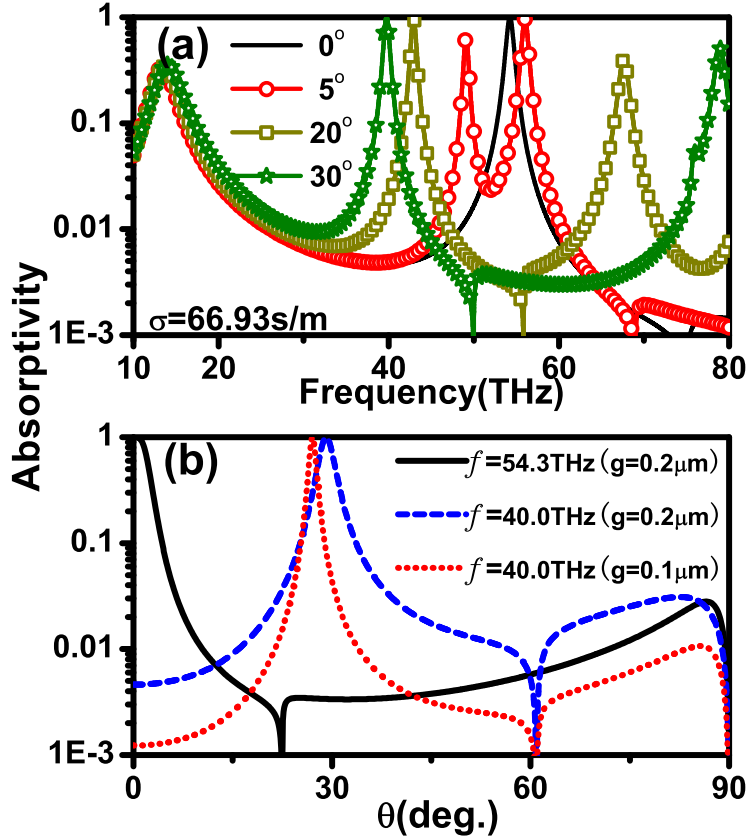}\caption{Absorption spectra under
TM-polarized incidence (a) as a function of frequency at incident angles of
$\theta=0^{\text{o}},5^{\text{o}},20^{\text{o}},30^{\text{o}}$ (g=0.2$\mu
$\textit{m}) and (b) as a function of incident angle at 54.3THz (solid line,
$g$=0.2$\mu$\textit{m}), 40\textit{THz} (dashed line, g=0.2$\mu$\textit{m})
and 40THz (dotted line, g=0.1$\mu$\textit{m})}%
\end{figure}

We derived the absorption spectra of the slab$A(\vec{k}_{0},\omega
)=1-\sum\limits_{m}{Re(\frac{k_{z_{m}}^{I}}{k_{z_{0}}^{I}})|r_{m}(\vec{k}%
_{0},\omega)|^{2}}$ which includes the contributions from all Bloch orders of
reflected waves. As a consequence, $A(\vec{k}_{0},\omega)$ gives information
about the surface resonance states as well as the emissivity properties as
governed by Kirchhoff's law\cite{30} . We shall assume that the dielectric
spacer layer is slightly dissipative by assigning a complex
permittivity$\varepsilon_{III}=\varepsilon_{r}\varepsilon_{0}+i\sigma/\omega$
with $\varepsilon_{r}=2.2$ and $\sigma=66.93$S/m [$Im(\varepsilon
_{III})\approx10^{-2}\varepsilon_{r}\varepsilon_{0}$] in the calculated
frequency regime. In Fig. 2(a), we present the absorption spectra at various
incident angles. The spectra exhibit a low and broad peak at $13.2$THz which
is almost independent of the incident angle, while the other absorption peaks
at higher frequencies are narrow and sensitive to the incident angle with a
maximum absorption approaching 100{\%}. The slab thus acts as an all-angle
absorber at $13.2$THz (a similar result can be found in Ref. 31), but exhibits
sharp angle-selective absorption peaks at higher frequencies. Shown as solid
and dashed lines in Fig. 2(b), the sharp angular dependence of absorption
coefficients (note that the vertical axis is in log-scale) at $40.0$THz and
54.3THz implicitly implies the existence of spatially coherent surface
resonance states. The angle-dependent absorption peaks become lower and
disappear gradually with the increase of the material loss. This presents a
way to realize nearly perfect absorption with weakly absorptive materials by
coherent surface resonance states. The coherent length of a surface resonance
state can be estimated by the ratio of the wavelength $\lambda$ and the full
width at half maximum (FWHM) $\Delta\theta$ of the absorption peak\cite{32}.
For example, for the $\Gamma_{4}$ state at 54.3THz and $\vec{k}_{0_{\parallel
}}=0$, the angular FWHM of the corresponding absorption peak $\Delta
\theta=4.6^{\text{o}}$(from $\theta=-2.3^{\text{o}}$to $\theta=2.3^{\text{o}%
})$ gives rise to a coherent length $\lambda/\Delta\theta=68.5\mu
m\approx12.4\lambda$. The coherent length is about $220\lambda$ for the
surface resonance state at 50.22THz and $\vec{k}_{0_{\parallel}}\approx
0.02\pi/p$ with $\Delta\theta=0.26^{\text{o}}$(not shown in figure). The
angular FWHM is reduced if the gap size is smaller, as shown with the dashed
and dotted lines in Fig. 2(b) for $g=0.2\mu$m and $g=0.1\mu$m at 40THz, which
means that the coherent length of the surface resonant modes can be controlled
by the gap-period ratio $g/p$.

\begin{figure}[ptb]
\centering
\includegraphics[width=12cm]{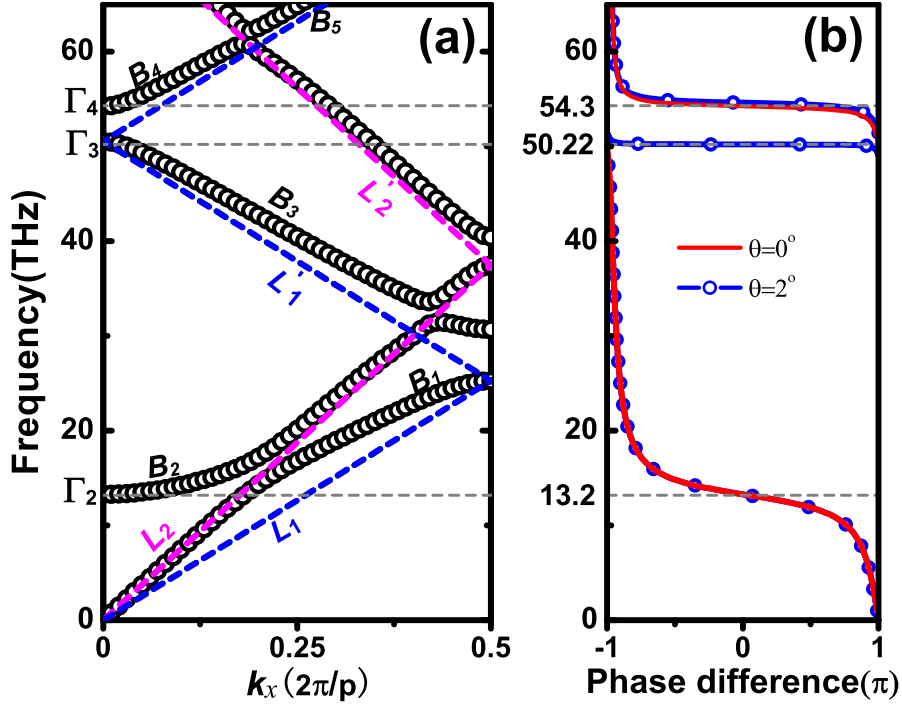}\caption{(a) Dispersion diagram of
TM-polarized surface resonance states. (b) The reflection phase difference
between the 0$^{th}$ order reflection and the TM-polarized incident plane wave
at incident angles of $\theta=0^{\text{o}}$(red line) and $\theta=2^{\text{o}%
}$(blue line). }%
\end{figure}

To quantitatively characterize the formation of these spatially coherent
surface resonance states, we employ the eigenmode expansion method to
calculate the surface resonance dispersion (in the limit of no material loss)
as shown in Fig. 3(a). The $B_{1}$ surface resonance states lie blow the light
line $L_{2}$ (magenta dashed line), and thus are non-radiative as evanescent
modes. The surface resonances labeled as $B_{2}$ originate from the coupling
of the fundamental magnetic resonance modes of the metal strip structure with
the free space light line $L_{2}$. The surface resonances $B_{3}$ and $B_{4}$
are harmonic modes of the magnetic resonances that hybridizes with the guided
mode inside the dielectric layer. The calculated reflection phase difference
between the 0$^{th}$ order reflected and incident electric field, as shown in
Fig. 3(b) for normal incidence(red line), and $2^{\text{o}}$ incidence(blue
line), clearly shows that the resonances are magnetic in nature when the
surface resonances intersect the zone center at $\Gamma_{2}$(13.2THz ) and
$\Gamma_{4}$(54.3THz) as the reflection phase is zero like what a magnetic
conductor surface does to the incident waves. The state $\Gamma_{3}$,
invisible in the reflection phase spectrum under normal incidence [red solid
line in Fig. 3(b)], is a dark state as its eigenmode is in mirror symmetry
about the $\hat{y}\hat{z}$ plane and can not couple with free space photons.
While the other $B_{3}$ states can couple with external light under oblique
incidence [see the blue line in Fig. 3(b)]. For example, there exists in-phase
reflection at frequency 50.22THz under an incident angle of 2$^{\text{o}}$,
corresponding a $B_{3}$ state at frequency 50.22THz and $k_{0_{\parallel}%
}\approx0.02\pi/p$.

\begin{figure}[ptb]
\centering
\includegraphics[width=15cm]{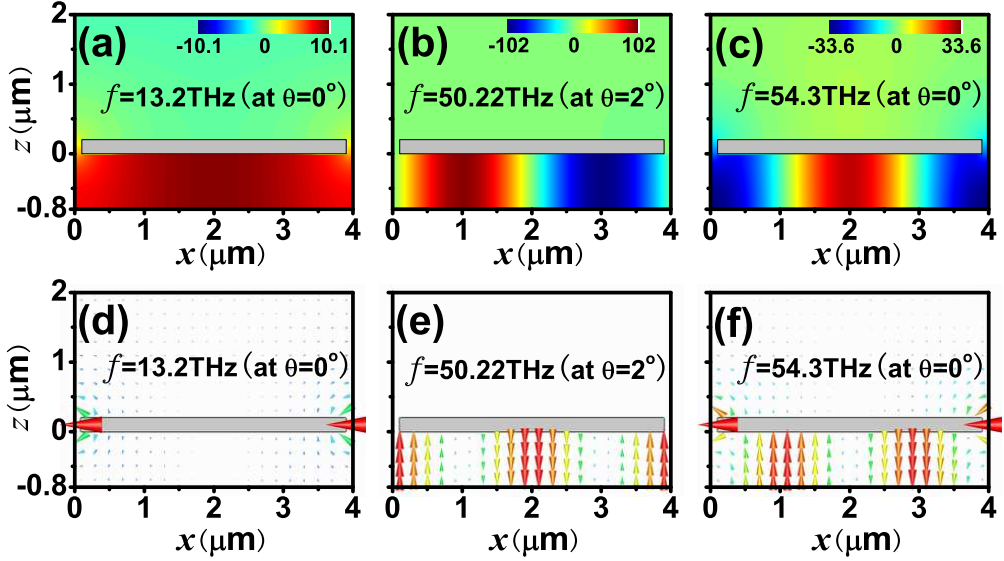}\caption{Spatial distributions of magnetic
fields and electric fields in the $\hat{x}\hat{z}$ plane for $\Gamma_{2}$
state at $f_{\Gamma_{2}}=13.2$THz, $\theta=0^{\text{o}}$[(a) and (c)], a state
on $B_{3}$ at $f=50.22$THz,$\theta=2^{\text{o}}$[(b) and (e)]and $\Gamma_{4}$
state at $f_{\Gamma_{4}}=54.3$THz, $\theta=0^{\text{o}}$[(c) and (f)]}%
\end{figure}

The angle-independent absorption peak at 13.2THz is due to the $B_{2}$ mode,
which is only weakly dispersive near the zone center. The more dispersive
$B_{3}$ and $B_{4}$ modes are accountable for the incident-angle sensitive
absorption in the higher frequencies in Fig. 2(a). The field patterns in Figs.
4(a)-4(c) present the spatial distribution of the real part of magnetic fields
excited by the incident plane waves with incident angles $0^{\text{o}}%
$,$2^{\text{o}}$ and $0^{\text{o}}$ for the three surface resonance states on
$B_{2}$, $B_{3}$ and $B_{4}$ respectively, and the corresponding vector
diagrams of electric fields are shown in Figs. 4(d)-4(f). We can see clearly
that the electric fields reach maximum in strength at the slab upper surface,
and exponentially decay along the surface normal into the free space. This is
precisely a picture of SPP modes. The field patterns comes from the
coincidence of the evanescent wave components in high Bloch orders at both
sides of metallic grating.

\begin{figure}[ptb]
\centering
\includegraphics[width=12cm]{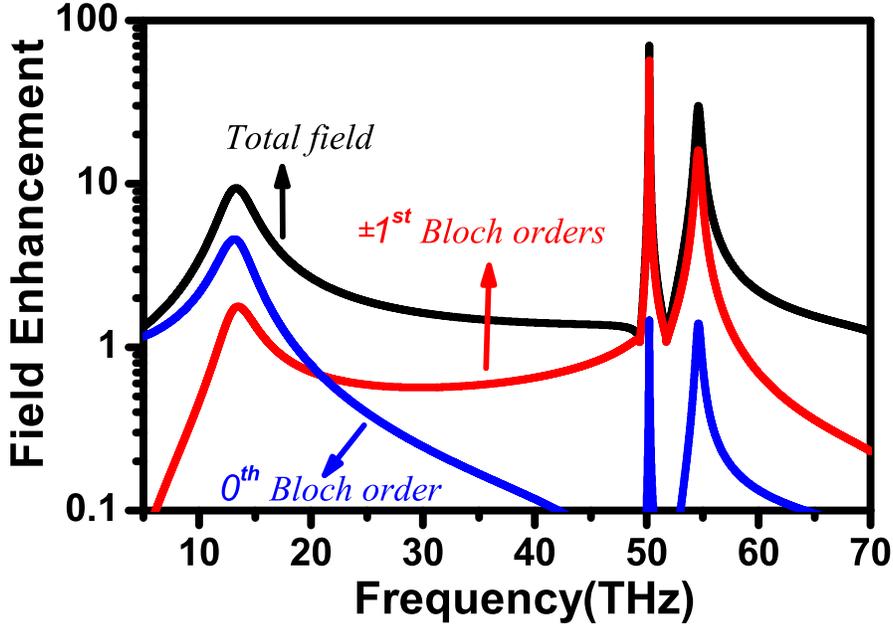}\caption{Magnetic field $\vec{H}$ inside the
dielectric layer normalized to that of incidence $H_{0}$ under an incident
angle of $\theta=2^{\text{o}}$. Black line: all Bloch orders of TEM guided
modes included; Red line: only the 0$^{th}$ Bloch order considered; Blue line:
summation of -1$^{st}$ and +1$^{st}$ Bloch orders of TEM guided modes.}%
\end{figure}

Although only TEM guided modes are allowed to be excited in the thin MDM slab
within the frequency of our interest, the $B_{2}$ states are quite different
from the $B_{3}$ and $B_{4}$ states in field patterns inside the dielectric
layer. We see from Figs. 4(b), 4(c) ,4(e)and 4(f) that for a $B_{3}$ or
$B_{4}$ state, there are nodes and anti-nodes in field patterns, while for the
$B_{2}$ state, the magnetic field is almost uniformly distributed.
Calculations on local field enhancement inside the dielectric slab resolve the
puzzle. Black solid line in Fig. 5 presents the normalized magnetic field
$|H|$ inside the dielectric with respect to that of incidence $|H_{0}|$ under
an incident angle of $\theta=2^{\text{o}}$. The Fourier component in $m=0$
order [blue solid line] contributes the most at 13.2THz and the least at
50.22THz and 54.3THz; while it is just opposite for the contributions in
combination from the two high order Fourier components with \textit{m=}$\pm1$
orders[red solid line]. Figure 5 also indicates that the enhancement of local
field of an excited $B_{3}$ or $B_{4}$\ state can be ten times larger than
that of an excited $B_{2}$ state; the enhancement factor at 50.22THz is about
100, while it is only 10 at 13.2THz.

We see from Fig. 3(a) that the surface resonance dispersion of the slab comes
from the interaction between the magnetic resonances and the (folded) light
lines $L_{1}$ (for dielectrics) and $L_{2}$ (for air) grazing on the
interfaces. In the limit of a small gap-period ratio ($g/p=0.05$ for example),
our system is weakly Bragg-scattered, and as such, when a surface resonance
state on branches $B_{3}$ or $B_{4}$ is excited, the induced wave fields
inside the dielectric of region III are guided quasi-TEM modes dominated by
$\pm1^{st}$ Bloch orders. For that reason, the $B_{3}$ and $B_{4}$ states have
high fidelity even though they are leaky modes, as most of their Bloch
wavefunction components lying outside the free space light line. As the air
gaps of the metallic grating serve to couple the electromagnetic waves of
region I and region III, the quality factor of a resonance state can be
estimated with the overlap integral between the fundamental waveguide mode
$|\alpha_{0}\rangle$ in the air gap and the dominant Bloch waves $|\vec{k}%
_{m}^{i}\rangle$ ($i=I,III)$ in region I or region III for the coupling
coefficients
\begin{equation}
C_{m}^{i}=\langle\alpha_{0}|\vec{r}|k_{_{_{m}}}^{i}\rangle=\frac{1}%
{\sqrt{\varepsilon_{r}}}\frac{k_{zm}^{i}}{\sqrt{k_{x}^{2}+k_{zm}^{i}{}^{2}}%
}\operatorname{sinc}[\frac{(k_{x}+G_{m})g}{2}], \label{eq3}%
\end{equation}
when the air gap width $g\ll p$ is satisfied. For the $B_{2}$ states, the
major Fourier component of the wavefunction is $|\vec{k}_{0}^{i}\rangle$ in
zero order, and as $k_{z0}^{III}$ is generally not small, $C_{0}^{III}$ is
usually very large according to Eq. 3, and the $B_{2}$ states leak out easily.
The states on branches $B_{3}$ and $B_{4}$ have major Fourier components in
\textit{m=}$\pm1$ order, and as they are asymptotic to the (folded) dielectric
light lines $L_{1}$, the absolute value of $k_{zm}^{III}$ ($m=+1$ for
$k_{x}<0$ or $m=-1$ for $k_{x}>0)$ is very small, resulting in the small
coupling coefficients $C_{-1}^{III}$ or $C_{+1}^{III}$. The $B_{3}$ and
$B_{4}$ modes have to travel a long distance before they leak out. They have a
long life time and a good spatial coherence. It also explains why the state
$\Gamma_{3}$, a state precisely superposing on folded light line
$L_{1}^{\prime}$ in dielectric layer, is dark to the incident plane wave as
$k_{zm}^{III}=0$.

Different from $B_{3}$ and $B_{4}$ states, the $B_{2}$ states have a major
Fourier component in $m=0$ order which directly couples to the free space
photons. As a consequence, the $B_{2}$ states, forming a flat band far away
from the light line $L_{2}$ when $\vec{k}_{0_{\parallel}}$ is small, are
localized with resonant frequency scaled by local geometry of unit cell. The
high model fidelity of a $B_{3}$ or $B_{4}$ state also gives rise to much more
intense local field compared to the $B_{2}$ states. As shown in Fig. 5, the
induced local field is 100 times stronger than the incident field for the
state on $B_{3}$; while it is only 10 times stronger for $\Gamma_{2}$, and
this is consistent with the absorptivity shown in Fig. 2(a). In addition, the
coherent length can be adjusted by the gap width as the kernel $\langle
\alpha_{0}|\vec{r}|k_{_{_{m}}}^{i}\rangle$ is proportional to the gap-period
ratio $g/p$. More calculations demonstrate that the angular FWHM of the
absorption peak is reduced from $0.26^{\text{o}}$ to $0.16^{\text{o}}$ when
the gap is decreased from $0.2\mu$m to $0.1\mu$m, corresponding to a coherent
length of $358\lambda$.

We note that most of the attentions in previous studies have been devoted on
the localized $B_{2}$ states\cite{22,23,24,31} . While the spatially coherent
surface resonance states will lead us into a new vision about coherent control
of emission radiations. J.-J. Greffet and co-workers showed that highly
directional and spatially coherent thermal emission can be obtained by etching
a periodic grating structure into a SiC surface\cite{32,33,34,35}. The
magnetic resonant modes in our system can do the same, as will be demonstrated
below. Our system has the advantage that the operational frequency is tunable
by changing the structural parameters, and the operational bandwidth is wide.
In addition, our structure supports all-angle functionality for some specific
range of frequencies as shown in Fig. 6, although it is periodic only in one direction.

\section{Coherent Thermal Emission}

\begin{figure}[ptb]
\centering
\includegraphics[width=14cm]{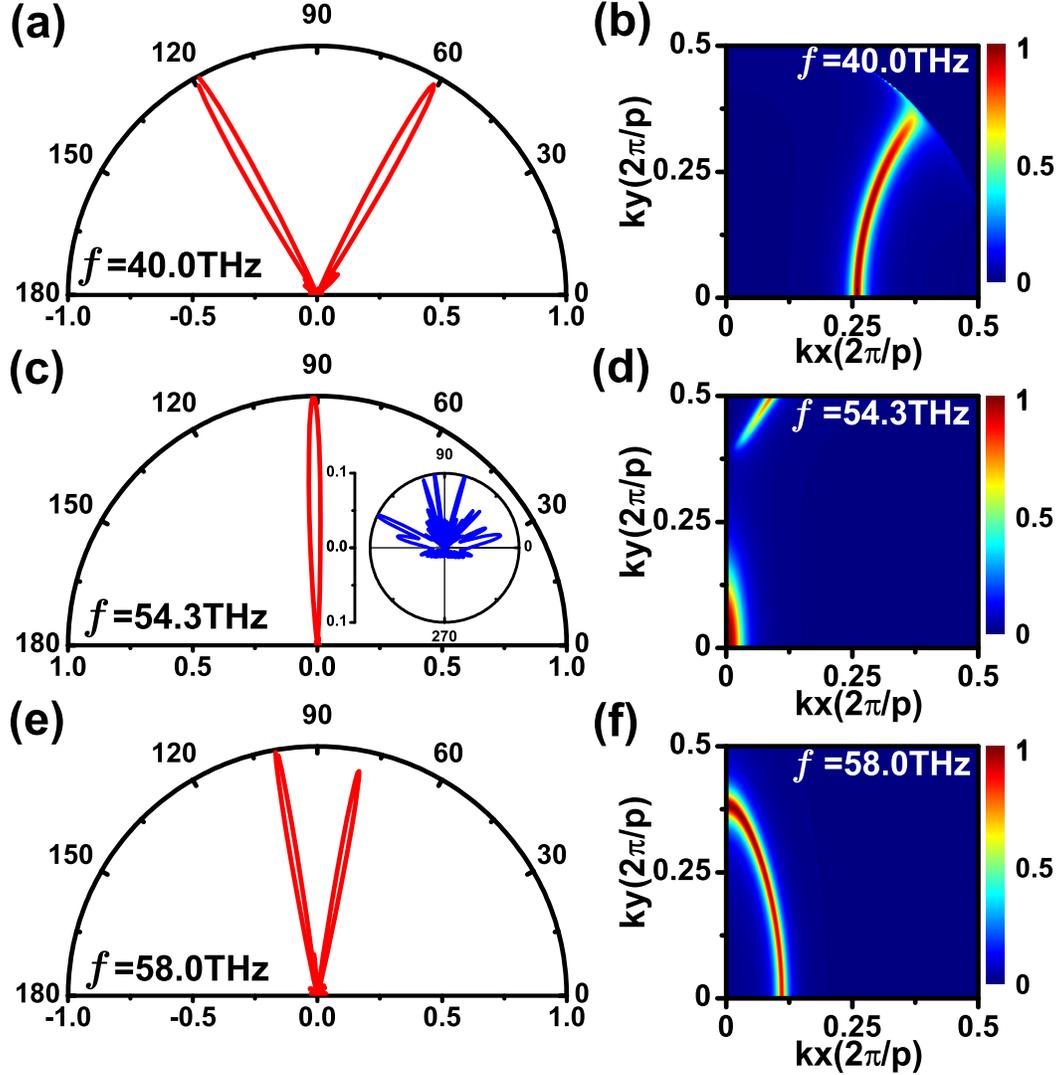}\caption{Radiation patterns in the H-plane
(calculated by FDTD) and absorptivity (calculated by mode expansion method) as
a function of in-plane wavevectors at $f=40.0$THz[(a) and (b)], $f=54.3$%
THz[(c) and (d)] and $f=58.0$THz[(e) and (f)]. In FDTD simulations, 1200 point
sources with random phases are placed at the mesh points inside the dielectric
layer. A 4{\%} structural disorder is included in the$80\mu$msimulation cell,
which accounts for the slight asymmetry of the radiation patterns, but also
demonstrates the robustness of the angle selectivity with respect to
disorder.The inset in (c) is a control calculation in which the top metal
gratings are removed, so that there is just a dielectric layer with random
phase sources above a metal ground plane. The directivity of emission from the
random sources is lost.}%
\end{figure}We performed finite-difference-in-time-domain (FDTD) simulations
to emulate the emissions from a slab containing point sources with random
phases using the same configuration parameters aforementioned. We purposely
put disorder in structure to test the robustness of the phenomena. We assigned
two Gaussian distributions(they can be uniform distributions or other types as
well) independently to the width of metallic strips and the center positions
of air gaps to introduce a 4{\%} (standard deviation) structural disorder. The
slab has a lateral size of 60 periods along the $\hat{x}$ direction. A total
of 1200 point sources with random phases are placed at the mesh points inside
the dielectric layer. Directional emissions of a wide range of frequencies
above $34$THz are confirmed by the simulation. The 4{\%} structural disorder
has little impact on the directional emissivity. Figs. 6 (a), 6(c) and 6(e)
show the far-field emission patterns in the $\hat{x}\hat{z}$ plane (H-plane)
at 40.0THz, 54.3THz and 58.0THz. The inset in Fig. 6(c) is a control
calculation in which the top metal gratings are removed, so that there is just
a dielectric layer with random phase sources above a metal ground plane. The
directivity of emission from the random sources is lost. Figs. 6(b), 6(d) and
6(f) present the absorptivity (under plane wave incidence) as a function of
in-plane wave-vector (solid angle) at these frequencies. The strong angle
selectivity of the absorption is evident, and by Kirchhoff's law, the thermal
emission should also be highly directional, which is a direct consequence of
the good spatial coherence of the surface resonance states. As shown in Figs.
6(b), 6(d) and 6(e), the absorption/emission peaks generally trace out an arc
in the $k_{x}\sim k_{y}$ plane, but near $54.3$THz[Fig. 6(d)], the dominant
emission beam is restricted to a small region near the zone center. This is
because the $\Gamma_{4}$ state is a minimum point if we consider the band
structure in the $k_{x}-k_{y}$ plane. That means that at 54.3THz, we can
obtain a directional emission beam not just in the H-plane, but in all
directions, although the structure is periodic in only one direction.

We note that there are other schemes to realize coherent thermal radiations,
such as by utilizing three-dimensional photonic crystals\cite{36} or
one-dimensional photonic crystal cavities\cite{37}. Our metamaterial slab
presents a route to achieve linearly polarized coherent thermal emission
radiations in a wide frequency range which can be tuned by adjusting
structural parameters and material parameters.

\section{Conclusion}

In summary, we proposed a simple metamaterial slab structure that possesses
spatially coherent magnetic surface resonance states in a broad range of
frequencies. These states facilitate nearly perfect absorption in a thin
metamaterial slab containing slightly absorptive materials. As the absorption
spectrum is highly angle-selective, the slab should give directional thermal
emission. Direct FDTD simulation with random-phase sources corroborates the
existence of strong angular emissivity even in the presence of structural
disorder. As the surface resonances originate from artificial resonators, the
operational frequency and the response can be tuned by varying the structural
configurations. The simple metamaterial structure may be a useful platform to
realize the coherent control of thermal emissions, optical antennas, infrared
or THz spectroscopy as well as photon detector.

\section{Acknowledgment}

This work is supported by the National 863 Program of China (Grant No.
2006AA03Z407), NSFC (Grant No. 10974144, No. 60674778), CNKBRSF (Grant
No.2006CB921701), HK RGC grant 600308, NECT, STCSM and Shanghai Education and
Development Foundation (No. 06SG24).


\begin{thebibliography}{99}                                                                                               %


\bibitem {1}Raether H 1988 \textit{Surface Plasmons on Smooth and Rough
Surfaces and on Gratings} (Berlin: Springer-Verlag)

\bibitem {2}Barnes W L, Dereux A and Ebbesen T W 2003 \textit{Nature}
\textbf{424} 824-30

\bibitem {3}Kamli A, Moiseev S A and Sanders B C 2008 \textit{Phys. Rev.
Lett.} \textbf{101} 263601

\bibitem {4}Vasa P, Pomraenke R, Schwieger S, Mazur Y I, Kunets V, Srinivasan
P, Johnson E, Kihm J E, Kim D S, Runge E, Salamo G and Lienau C 2008
\textit{Phys. Rev. Lett.} \textbf{101} 116801

\bibitem {5}Andrew P, Kitson S C and Barnes W L 1997 \textit{J. Mod. Opt.}
\textbf{44} 395-406

\bibitem {6}Coutaz J L, Neviere M, Pic E and Reinisch R 1985 \textit{Phys.
Rev. B} \textbf{32} 2227-32

\bibitem {7}Tsang T Y F 1996 \textit{Opt. Lett.} \textbf{21} 245-7

\bibitem {8}Kneipp K, Wang Y, Kneipp H, Perelman L T, Itzkan I, Dasari R R and
Feld M S 1997 \textit{Phys. Rev. Lett.} \textbf{78} 1667-70

\bibitem {9}Nie S and Emory S R 1997 \textit{Science} \textbf{275} 1102-6

\bibitem {10}Pendry J B, Holden A J, Stewart W J and Youngs I 1996
\textit{Phys. Rev. Lett.} \textbf{76} 4773-6

\bibitem {11}Pendry J B, Holden A J, Robbins D J and Stewart W J 1999
\textit{IEEE Trans. Microwave Theory Tech.} \textbf{47} 2075-84

\bibitem {12}Sievenpiper D, Zhang L J, Broas R F J, Alexopolous N G and
Yablonovitch E 1999 \textit{IEEE Trans. Microwave Theory Tech.} \textbf{47} 2059-74

\bibitem {13}Pendry J B, Martin-Moreno L and Garcia-Vidal F J 2004
\textit{Science} \textbf{305} 847-8

\bibitem {14}Hibbins A P, Evans B R and Sambles J R 2005 \textit{Science}
\textbf{308} 670-2

\bibitem {15}Liu H, Genov D, Wu D, Liu Y, Steele J, Sun C, Zhu S and Zhang X
2006 \textit{Phys. Rev. Lett.} \textbf{97} 243902

\bibitem {16}Lockyear M J, Hibbins A P and Sambles J R 2009 \textit{Phys. Rev.
Lett.} \textbf{102} 073901

\bibitem {17}Grigorenko A N, Geim A K, Gleeson H F, Zhang Y, Firsov A A,
Khrushchev I Y and Petrovic J 2005 \textit{Nature} \textbf{438} 335-8

\bibitem {18}Shalaev V M 2007 \textit{Nat. Photon.} \textbf{1} 41-8

\bibitem {19}Boltasseva A and Shalaev V M 2008 \textit{Metamaterials}
\textbf{2} 1-17

\bibitem {20}Zhou L, Wen W, Chan C and Sheng P 2003 \textit{Appl. Phys. Lett.}
\textbf{83} 3257-9

\bibitem {21}Engheta N and Ziolkowski R W 2006 \textit{Metamaterials: physics
and engineering explorations}: Wiley {\&} Sons.)

\bibitem {22}Hibbins A, Sambles J, Lawrence C and Brown J 2004 \textit{Phys.
Rev. Lett.} \textbf{92} 143904

\bibitem {23}Hibbins A, Murray W, Tyler J, Wedge S, Barnes W and Sambles J
2006 \textit{Phys. Rev. B} \textbf{74} 073408

\bibitem {24}Brown J, Hibbins A, Lockyear M, Lawrence C and Sambles J 2008
\textit{J. Appl. Phys.} \textbf{104} 043105

\bibitem {25}Sheng P, Stepleman R S and Sanda P N 1982 \textit{Phys. Rev. B}
\textbf{26} 2907-16

\bibitem {26}Lalanne P, Hugonin J P, Astilean S, Palamaru M and Moller K D
2000 \textit{J. Opt. a-Pure Appl. Op.} \textbf{2} 48-51

\bibitem {27}Wei Z, Fu J, Cao Y, Wu C and Li H 2010 \textit{Photonics and
Nanostructures - Fundamentals and Applications} \textbf{8} 94-101

\bibitem {28}Wei Z, Li H, Wu C, Cao Y, Ren J, Hang Z, Chen H, Zhang D and Chan
C T 2010 \textit{Opt. Express} \textbf{18} 12119-26

\bibitem {29}Jackson J D 1998 \textit{Classical Electrodynamics} (New York: Wiley)

\bibitem {30}Greffet J-J and Nieto-Vesperinas M 1998 \textit{J. Opt. Soc. Am.
A} \textbf{15} 2735-44

\bibitem {31}Diem M, Koschny T and Soukoulis C 2009 \textit{Phys. Rev. B}
\textbf{79} 033101

\bibitem {32}Greffet J-J, Carminati R, Joulain K, Mulet J-P, Mainguy S and
Chen Y 2002 \textit{Nature} \textbf{416} 61-4

\bibitem {33}Le Gall J, Olivier M and Greffet J-J 1997 \textit{Phys. Rev. B}
\textbf{55} 10105-14

\bibitem {34}Carminati R and Greffet J-J 1999 \textit{Phys. Rev. Lett.}
\textbf{82} 1660-3

\bibitem {35}Marquier F, Joulain K, Mulet J-P, Carminati R, Greffet J-J and
Chen Y 2004 \textit{Phys. Rev. B} \textbf{69} 155412

\bibitem {36}Laroche M, Carminati R and Greffet J-J 2006 \textit{Phys. Rev.
Lett.} \textbf{96} 123903

\bibitem {37}Lee B J, Fu C J and Zhang Z M 2005 \textit{Appl. Phys. Lett.}
\textbf{87} 071904
\end{thebibliography}
\end{document}